\documentclass[useAMS,usenatbib]{mn2e}
\usepackage{amsmath}
\usepackage{float}
\usepackage{listings}
\usepackage{graphicx}
\usepackage{natbib}
\usepackage{multirow}

\title{Cosmic flows in the nearby universe from Type Ia Supernovae}
\author[S.J. Turnbull et al.]
{Stephen J. Turnbull$^1$, Michael J. Hudson$^{1,2}$, Hume A. Feldman$^3$, Malcolm Hicken$^4$, 
\newauthor
Robert P. Kirshner$^4$, Richard Watkins$^5$ \\
$^1$ Department of Physics and Astronomy, University of Waterloo,
Waterloo, Ontario, N2L 3G1, Canada\\
$^2$ Perimeter Institute for Theoretical Physics, 31 Caroline St N,
Waterloo, Ontario, N2L 2Y5, Canada\\
$^3$ Department of Physics \& Astronomy, University of Kansas, Lawrence, KS 66045, USA\\
$^4$ Harvard-Smithsonian Center for Astrophysics, 60 Garden Street, Cambridge, MA 02138 USA\\ 
$^5$ Department of Physics, Willamette University, Salem, OR 97301, USA\\
E-mail: s4turnbu@uwaterloo.ca\\
}

\begin{document}

\input{Step.sty}
\date{\today}
\maketitle
\label{firstpage}

\begin{abstract}
Peculiar velocities are one of the only probes of very large-scale mass density fluctuations in the nearby Universe.  We present new ``minimal variance'' bulk flow measurements based upon the ``First Amendment'' compilation of 245 Type Ia supernovae (SNe) peculiar velocities and find a bulk flow of 249 $\pm$ 76  $\kms$ in the direction \lberr{319}{18}{7}{14}. The SNe bulk flow is consistent with the expectations of $\Lambda$CDM. However, it is also marginally consistent with the bulk flow of a larger compilation of non-SNe peculiar velocities \citep*{WatFelHud09}. By comparing the SNe peculiar velocities to predictions of the IRAS Point Source Catalog Redshift survey (PSCz) galaxy density field, we find $\Omega_{m}^{0.55}\sigma_{8,\mathrm{lin}}$ = 0.40 $\pm$ 0.07, which is in agreement with $\Lambda$CDM. However, we also show that the PSCz density field fails to account for $150\pm43 \kms$ of the SNe bulk motion.
\end{abstract}

\section{Introduction}

In the standard cosmological model, gravitational instability causes the growth of structure and peculiar velocities. In the regime where the perturbations are linear, there is a simple relationship between density and peculiar velocity \citep{PPC}:
\begin{equation}
\mathbf{v}(\mathbf{r}) = \frac{f}{4 \pi} \int_{0}^{\infty} d^3\mathbf{r}'\delta(\mathbf{r}') \frac{\mathbf{r}'-\mathbf{r}}{\mid \mathbf{r}'-\mathbf{r}\mid^{3}}
 \label{eq:vpred}
\end{equation} 
where the growth factor $f$ is equal to $\Omega_{m}^{0.55}$ in flat $\Lambda$CDM models \citep{Lin05}, $\delta$ is the normalized mass density fluctuation field, $\delta = (\rho - \bar{\rho})/\bar{\rho}$, and $\mathbf{r}$ are coordinates in units of \kms.  

Given set of peculiar velocities, one can define a bulk flow as their average velocity; ideally the peculiar velocity tracers are dense and numerous enough that the resulting average is representative of the velocity of the volume. The bulk flow is then primarily due to structures on scales larger than the volume over which the bulk flow is measured (see Appendix A of \citealt{JusVitWys90} for a derivation).  Hence, bulk flows are probes of the large-scale power spectrum of \emph{matter} density fluctuations.

The $\Lambda$CDM  model, once normalized by WMAP7 \citep{LarDunHin11} observations of the Cosmic Microwave Background (CMB), fully specifies the  r.m.s.\   fluctuations of $\delta$ on all scales, and hence the cosmic r.m.s.\  of bulk flows \citep{WatFelHud09}.  While most studies of bulk flows agree on the general direction of the flow,  there is some disagreement as to the amplitude and scale.  \citet{WatFelHud09} applied a ``Minimal Variance'' (MV) weighting scheme to a compilation of 4481 peculiar velocity measurements.  Their results correspond to a sample with an effective Gaussian window of 50 $\hmpc$ and show a bulk flow of 407  $\pm$ 81 $\kms$ towards \lberr{287}{9}{8}{6}, which is in conflict with $\Lambda$CDM + WMAP7 at the 98 percent confidence level. The most controversial bulk flow result is the kinetic Sunyaev-Zeldovich flow dipole reported by \citet{KasAtrEbe10}, who found a bulk flow on the order of 1000 $\kms$ in the direction of \lberr{296}{28}{39}{14} over a scale of at least 800 \hmpc. If correct, this result would strongly conflict with $\Lambda$CDM + WMAP7.  

Another approach to understanding large-scale motions is to try to reconstruct the motion of the LG with respect to the CMB  \citep[627 $\pm$ 22 $\kms$ towards \lberr{276}{3}{30}{2};][]{KogLinSmo93} by measuring the distribution of galaxies and calculating the peculiar velocity of the Local Group (LG) using Eq. \eqref{eq:vpred}. Given the gravitational instability model of linear theory, the predicted velocity should converge to the measured CMB dipole for a sufficiently large survey volume. The application of Eq. \eqref{eq:vpred} is difficult in practice  because there are few redshift surveys that are both all-sky and deep. For example, \citet{ RowShaOli00} found that the predicted dipole from the IRAS Point Source Catalog Redshift survey \citep[][hereafter PSCz]{SauSutMad00} converged to 13.4 degrees of the CMB dipole by 30,000 $\kms$. However, \citet{BasPli06} reanalyzed the same data set and found that significant power was required on large scales, which was missed by the original analysis. Other studies have been based on the Two Micron All-Sky Survey Redshift Survey \citep[hereafter 2MRS]{SkrCutSti06}: \cite{ErdHucLah06} and found probable convergence, but \citet{LavTulMoh10} concluded that convergence was not obtained by 12,000 $\kms$, and may not be until well beyond 20,000 $\kms$.  In another study, using only the infrared fluxes,  \citet{BilChoMam11} concluded that even at an effective distance $\sim 300 \hmpc$ ($K_{s} < 13.5$) the flux dipole had not converged. 

In this paper we use Type Ia SNe as our peculiar velocity tracers. SNe have also been used as peculiar velocity probes by a number of authors \citep{RiePreKir95,RieDavBak97,RadLucHud04,PikHud05,LucRadHud05,HauHanTho06,ColMohSar11,DaiKinSto11}. 

An outline of this paper is as follows: In section 2, we introduce the data sets that were used. Section 3 presents the bulk flow of the SNe, using both simple weighting schemes as well as the ``Minimal Variance'' scheme of \citet{WatFelHud09}. Section 4 compares individual SNe peculiar velocities to the predictions of the IRAS PSCz density field. We discuss the implications of our results in Section 5, and present our conclusions in Section 6. Throughout, we adopt $\Omega_{m}=0.3$, $\Omega_{\Lambda}=0.7$, and quote distances in units of \kms.

\section{Data and Calibration}

In this study, three primary data sets of nearby SNe (with distances less than 20,000$\kms$) are combined. 

We refer to the first of these data sets as the `Old' sample and it contains 106 SN the youngest being from 2002, drawn from two sources: \citet{JhaRieKir07} and \citet{HicWooBlo09}.   Of the SNe in the `Old' sample, 34  are from \citet{JhaRieKir07}. The remaining 72 SN in `Old' are from \citet{HicWooBlo09}. The second data set, which we refer to as `Hicken', contains the remaining 113 SNe from \citet{HicWooBlo09} after cutting objects at distances larger than 20,000 $\kms$ and cutting two more objects (sn2007bz and sn2007ba) because they deviated by more than 3$\sigma$ after the first round of fitting (as described below). The last set is the recently released data set from `The Carnegie Supernova Project' \citep[][hereafter CSP]{FolPhiBur10}, containing 28 SNe.  Two of these objects were discarded due to our 20,000 $\kms$ distance cut, leaving 26 usable SNe. The CSP's reported uncertainties only reflected the derived distance modulus residual spread. A second intrinsic uncertainty($\sigma_{SN} $) in the magnitude of the SNe was added in quadrature by fitting a flow model and reducing the reduced chi-squared fit to 1.00. The intrinsic uncertainty was found to be 0.107 mag (slightly smaller than the 0.12 mag found by the CSP due to cuts and the additional free parameters of bulk flow). For further discussion of the light curve fitting, and consequences there of, for the `Old' and `Hicken' data sets see Appendix A.

	We combine these three sets to create a new sample that we dub the `First Amendment' (A1) compilation which we consider to be an extension to the  `Constitution' data set\footnote{The 'Old' and 'Hicken' sets combined  resemble very closely the `Constitution' set from \citet{HicWooBlo09} in terms of which supernovae are included. The light curve fitter used here (MLCS2k2) differs from that of the `Constitution' data set (SALT2).}. 

Where the observed SNe in the data sets were known to be contained within a cluster of galaxies, the redshift of the cluster was used for the observed velocity distance rather than the redshift of the supernova itself. Substituting cluster velocities for supernova velocities removes a significant source of thermal noise as objects in clusters can have a velocity rms of thousands of $\kms$ .  This process was applied to all three data sets. For galaxies not in clusters, the redshift of the host galaxy was used if the host galaxy redshift was recorded in NED, which occurred in all but two cases. For the remaining two cases we used the redshifts of the SNe. Galactic longitudes and latitudes for the Carnegie set were also taken from NED. 

The A1 data set has a characteristic or uncertainty-weighted depth of 58 \hmpc , where we define the characteristic depth to be: 
\begin{equation}
r_{*} = \frac{\Sigma\ r/ \sigma^{2}}{\Sigma\ 1/ \sigma^{2}} 
\end{equation}
where $\sigma$ is the total uncertainty in each SNe's peculiar velocity and $\mathbf{r}$ is the coordinates in units of \kms . 

\begin{figure*}
\begin{center}
\includegraphics[width=\textwidth]{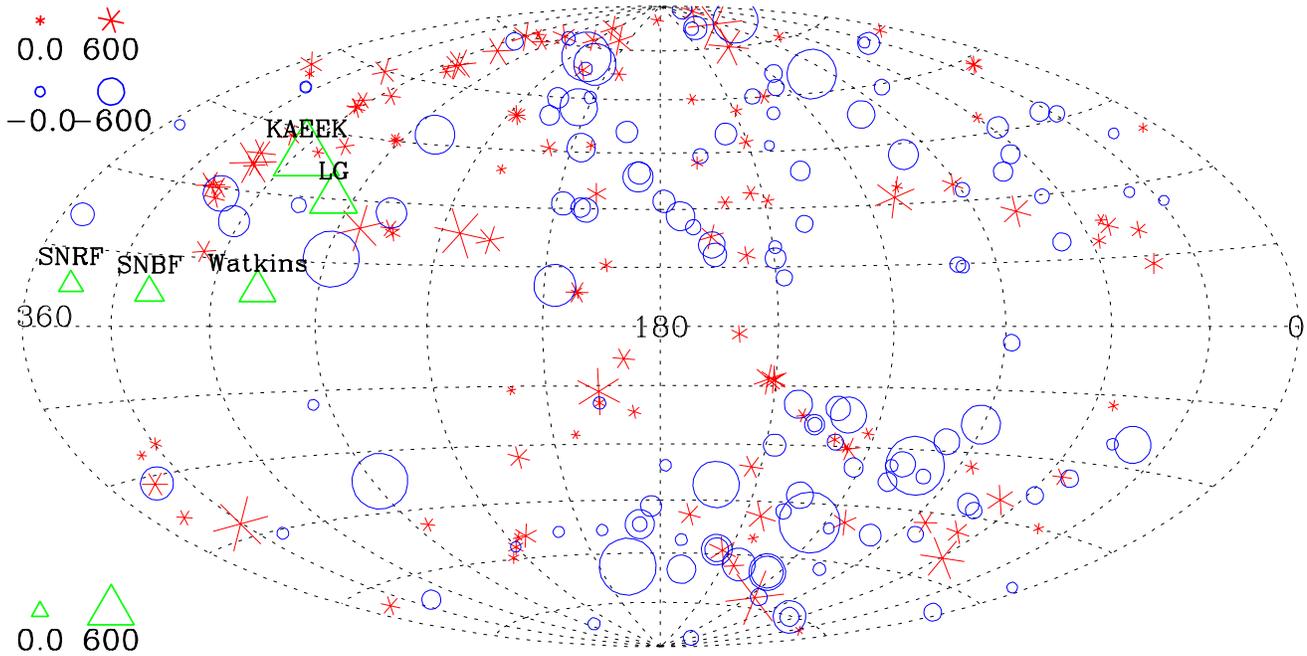}
\end{center}
\caption{An Aitoff projection of our data with circles (asterisks) representing SNe with peculiar velocities towards (away from) the LG. Larger symbols represent larger peculiar velocities in accordance with the scale shown top and bottom left. Also plotted in triangles are the direction motion of the LG with respect to the CMB, the \citet{WatFelHud09} bulk flow direction, the kinetic Sunyaev-Zel'dovich bulk flow direction of \citet[labelled KAEEK]{KasAtrEbe10}, and our new results (labeled SNBF for the bulk flow results from Section 3 and SNRF for the residual flow discussed in Section 4).}\label{fig:Aitoff}
\end{figure*}

\begin{figure}
\begin{center}
\includegraphics[width=\columnwidth]{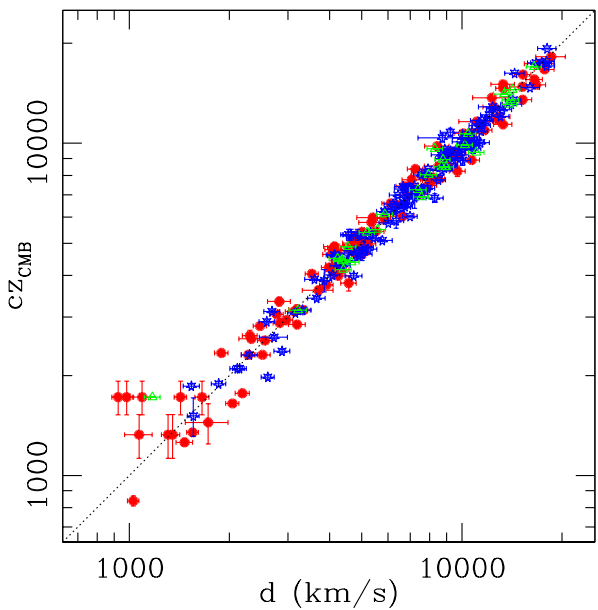}
\end{center}
\caption{A Hubble diagram showing the three subsets that make up the A1 data set:  `Old'  (red filled circles) , `Hicken' (Blue stars) and `Carnegie'(green triangles). The error bars can be seen to be approximately constant in the log log diagram, or increasing proportionally to the distance, as to be expected with the dominant error in most cases being the intrinsic uncertainty in SNe.}\label{fig:AHubbub}
\end{figure}

In Fig. \ref{fig:Aitoff} we present our results, our raw data, and the bulk flow directions that other surveys have found in an Aitoff projection. In Fig.  \ref{fig:AHubbub} we present the A1 data set in a Hubble Diagram divided into its three subsets. For all three data sets the intrinsic uncertainty of SNe's is the dominant source of error. Thus, for all our SNe the  percent error is approximately 6 percent of the measured distance, with the scatter for the `Old' and `Hicken' subsets being larger. 

\section{Bulk Flow}

In this section we discuss the bulk flow, which is the simplest statistic that can be derived from a peculiar velocity survey. 

\subsection{Methods}

We use two methods to measure the bulk flow. The first is a Maximum Likelihood (ML) method that minimizes the measurement uncertainties. The ML method is the traditional method used and  we consider it in order to compare new results with previous ML results.  However, ML methods have the disadvantage of returning the bulk flow of a specific sparse \emph{sample} of peculiar velocity tracers rather than the bulk flow of a regular \emph{volume}.  Comparisons between ML results are complicated by the different spatial sampling.  Instead, what is of greater interest is the bulk flow of a standardized volume. To estimate this, we calculate the ``minimum variance'' (MV) bulk flow as first introduced by \cite{WatFelHud09}.

\subsubsection{Maximum Likelihood}\label{sec:MaxL}

In general, we fit a simple flow model ($v_{\textrm pred}$) to the SNIa peculiar velocity data. In the case of the bulk flow $\mathbf{V}$ in the CMB frame, this flow model reduces to the radial component of the bulk flow vector for each SNe, i.\  e.:
\begin{equation}
 v_{\mathrm{pred},i} =\mathbf{V} \cdot \mathbf{\hat{r}}_{i}
\end{equation} 
where $\hat{r}_{i}$ is the unit vector pointing to each supernova. 

In the maximum likelihood method the weights are simply determined by the total uncertainty on the peculiar velocity of each object. Uncertainties in the observed peculiar velocity can be approximated well by a Gaussian,  in which case the Maximum-Likelihood solution can be approximated by  the following  $\chisq$:
\begin{equation}
 \chisq = \sum_{i} \frac{[cz_{obs,i} - (r_{i} + v_{\mathrm{pred},i} ) ]^{2}}{\sigma_{i}^{2}}
 \label{eq:Chi2}
\end{equation}
where $cz_{obs}$ is the observed redshift in $\kms$, $r_{i}$ is the distance converted from the reported distance modulus in $\kms$, $v_{\mathrm{pred},i}$ is the model velocity we are trying to measure as predicted for SNe {\it i}, and  $\sigma_{i}$ is the total uncertainty on the peculiar velocity of object {\it i} in units of \kms. This total uncertainty is the quadrature sum of the measurement error $\sigma_{m,i}^{2}$, the intrinsic uncertainty on SNe magnitude $\sigma_{SN}$ (both converted from magnitudes to \kms) and a ``thermal noise'' term ($\sigma\sbr{th}$) in units of \kms\ due to uncertainties in the flow model, such that:
\begin{equation}
 \sigma_{i}^{2} =\sigma_{m,i}^{2}+\sigma_{SN}^{2}+\sigma_{th}^{2}
\end{equation}
Note that since the $\sigma_{SN}^{2}$ term is converted from an uncertainty on magnitude, it is proportional to the distance to the SN. $\sigma_{SN}^{2}$ is often the dominant source of uncertainty since the thermal term is only important in nearby supernovae where the $\sigma_{m,i}^{2}$ and $\sigma_{SN}^{2}$ terms are small. The results are only weakly dependant upon the precise value chosen. Here, where the flow model is a simple bulk flow, we set the thermal noise to 250 $\kms$, which is consistent with previous work. The impact of this choice for the thermal noise is discussed in the results below.

We let each sub-sample of the A1 data set have a freely-varying independent Hubble term to identify degeneracies, to avoid underestimating final uncertainties, and to account for the fact that each subset may have slightly different calibration. None of the fits preferred a Hubble value that varied by more than  1 percent from the value drawn from the original sources. 

\subsubsection{Minimum Variance Method}

While the ML method described above is the best estimator of the bulk flow of a sparse sample, it is restricted in that it can only really characterize a particular survey, that will have its own errors and a specific and somewhat ill-defined geometry. The ML method is also, in a sense, density sampled, with higher density regions being more likely to contain a SN than voids. Most importantly, because the weights in the ML method are determined by the uncertainty on position in $\kms$, ML methods can be dominated by nearby SN that have smaller distance uncertainties. 

To better approximate a \emph{volume}-weighted bulk flow, we use the prescription described in \citet{WatFelHud09} to estimate the volume flow. Each SN is weighted so as to minimize the variance between the bulk flow measured in the real sample and the bulk flow as it would be measured in a perfectly-sampled 3D Gaussian.  We adopt a Gaussian with an ``ideal'' radius $R_I = 50 \hmpc$. Effectively, weights are assigned to each SN based on their proximity to other SNe in the data set, and on how they compare with an ideal uniform sampling. This weighting scheme is specifically designed to maximize sensitivity to large scales. The MV weighting scheme has been tested using mock catalogues drawn from N-body simulations by Agarwal et al. 2011 (in preparation), who demonstrate that the recovered MV bulk flows are unbiased and have errors within the range expected from linear theory.

\subsection{Consistency of SNe Subsamples}

Before analysis of the combined SNe sample is undertaken, it is important to confirm that the data subsamples agree with one another.  We calculate a $\chi^{2}$ statistic for each pair of subsamples, following the analysis of \citet{WatFelHud09}, which accounts for sparse sampling effects. The $\chi^{2}$ statistic we use is given by the equation,
\begin{equation}
\chi^2 = \sum_{i,j} (\Delta V_{i})(C_{i,j})^{-1}(\Delta V_{j}).
\label{eq:chisq}
\end{equation}
where $\Delta \vec{V}$ is the bulk flow vector, and \textbf{C} is the covariance matrix taking into account the window functions of both surveys and the power spectrum, (see  in equations 21 - 23 of \citet{WatFelHud09}). The results are shown in  Table \ref{tab:consist}. In summary, we find that all three subsamples are consistent with each other.

\begin{table}
\caption{$\chi^2$ for 3 DoF for the surveys for $\Omega_{m}=0.258$. If the $\chi^{2}$ value is greater than 7.8, the two surveys disagree at a greater than 95\% confidence level. The probabilities resported are the $\chi^{2}$ test probability of agreement between the two.} 
\begin{tabular}{lccc}
\hline \hline
&   $R_I=$ & \multicolumn{2}{c}{50\hmpc} \\
Survey             &   & $\chi^{2}$&  Probability \\  \hline \hline 
Old vs Hicken      &   & 0.173  &  98.2  \\ \hline
Old vs Carnegie    &   & 2.293  &  51.4  \\ \hline
Hicken vs Carnegie &   & 1.369  &  71.3  \\ \hline
\hline
\label{tab:consist}
\end{tabular} 
\end{table}

\subsection{Results}

In Table \ref{tab:SimpMod}, we present a summary of the results from the bulk flow, subdivided by data set and by weighting scheme.  The ML bulk flow for the A1 sample was found to be  197 $\pm$ 56  $\kms$  in direction \lberr{295}{16}{11}{14}. This is significantly different from zero at the 99.9\% confidence level. 

As discussed above, the ML method gives most weight to SNe with the lowest errors in units of \kms, i.e. the nearest SNe. In order to reduce the impact of these nearby SNe, it is interesting to redetermine the bulk flow excluding nearby objects.  The middle section of \tabref{SimpMod} shows the bulk flow using only SNe with $6000 \kms < d < 20000 \kms$. This subsample indicates a slightly higher amplitude flow, albeit with larger error bars: $330\pm120 \kms$ towards \lb{321}{20}.

Finally, the MV results shown in the third section of \tabref{SimpMod} should give the most robust estimates of the flow of a Gaussian volume of radius 50 \hmpc. For the entire A1 sample, the MV flow is 248 $\pm$ 87  $\kms$ in the direction \lberr{319}{25}{7}{13}.

These values are lower than the LG's motion in the CMB frame, indicating that some of the LG's motion must come from structures within our survey volume (such as the Virgo and Hydra-Centaurus superclusters).  

To investigate the sensitivity of our results to the value of the thermal noise we adjusted it by $\pm$ 100 $\kms$; When so tested, the final magnitude of the A1 sample MV flow only changed by $\pm$ 31 $\kms$.

\begin{table*}
\caption{ Bulk flow for all three SNIa data subsets and the combined First Amendment set. For comparison MV50 result from \citet{WatFelHud09} are also included. Note the uncertainties quoted for the ML method are the just propagated uncertainties from measurements, the uncertainties for the MV method also include an approximation of additional noise due to non-uniform sampling.}\label{tab:SimpMod}
\begin{tabular}{|l||c|         r@{$\pm$}l    |    r@{$\pm$}l    |    r@{$\pm$}l   |    r@{$\pm$}l    |     r@{$\pm$}l    |    r@{$\pm$}l    |} \hline\hline
               &Number&\multicolumn{2}{|c|}{Mag}&\multicolumn{2}{|c|}{l\arcdeg}&\multicolumn{2}{|c|}{b\arcdeg}&\multicolumn{2}{|c|}{V$_{X}$}&\multicolumn{2}{|c|}{V$_{Y}$}&\multicolumn{2}{|c|}{V$_{Z}$}         \\
\hline
\multicolumn{8}{|l|}{ML Thermal noise=250 \kms}&\multicolumn{2}{|c|}{\kms}&\multicolumn{2}{|c|}{\kms}&\multicolumn{2}{|c|}{\kms} \\
\hline
Old            &106       &226& 76&307& 21& 4&15 &136&76 &-180& 81&-35& 59\\			
Hicken         &113       &142&85 &283&41&30 &37 &27 &94 &-120&96 & 71& 78\\			                     
Carnegie       &26        &260&140&330&170&76 &38 &54 &182&-35 &230&250&150\\			 
First Amendment&245       &196&55 &300&17 &15 &14 &94 &55 &-165&58 &50 &44 \\ \hline     
 \hline
\multicolumn{14}{|l|}{ML Thermal noise=250 \kms, $d> 6,000 \kms$} \\
\hline
Old            &45        &450&190&331& 26&6& 21&390&200&-210&190&44&160\\		       
Hicken         & 76       &280&180&313&33 &27 &25 &170&170&-180&190&130&110\\		           
Carnegie       &15        &1132&850&117&14&16 &20&-490&540&970 &810&310&300\\	  
First Amendment&136       &330&120&321&20 &16 &15 &250&120&-200&130& 94&84\\ \hline	
\hline
\multicolumn{14}{|l|}{MV Weighting $R_{I}= 50 \hmpc$ Thermal noise=250 \kms} \\
\hline
Old            &113       &240&110&318&26 & -4&21 &180&110&-160&110&-16& 86\\		           
Hicken         &113       &250&110&310&25 &  5&20 &160&110&-190&110& 20& 85\\           
Carnegie       &28        &250&150&  0&340&81 &43 & 40&190&0   &240&250&150\\			  
\textbf{First Amendment} &\textbf{254}& \textbf{249}& \textbf{76}& \textbf{319} &\textbf{18} &\textbf{7}&\textbf{14} &\textbf{186}& \textbf{75} &\textbf{-162} &\textbf{77} &\textbf{32} &\textbf{59} \\ \hline
\citet{WatFelHud09}       &4481&407&81&287&9  &8  &6  &114&49&-387&53&57&37\\
\hline
\hline
\end{tabular}

\end{table*}

\subsection{Bulk Flow: Cosmology and Comparisons}

It is interesting to compare our ML bulk flow result to that of \citet{ColMohSar11}, who apply a maximum likelihood bulk flow fit to the Union2 catalogue of Type Ia SNe \citep{AmaLidRub10}. The Union2 catalogue contains 557 SNe, of which 165 are within 30,000 $\kms$. \citet{ColMohSar11}'s analysis yields a bulk flow velocity of 260 $\pm$ 150 $\kms$ based on SNe within 18000 $\kms$.  Our A1 sample yields a ML result of 196 $\pm$ 55 $\kms$, which is consistent with theirs. It must be noted that the agreement between these results is not as significant as might at first be assumed because there is significant overlap between the data sets.   However, Union2 uses SALT2 rather than MLCS2k2 ($R_{V}$=1.7) to obtain SN distances from the light-curve data.
 
\citet{DaiKinSto11} also analyzed the Union2 catalog, spliting it into two subsets. They defined a nearby set with 132 SNe at $z< 0.05$ for which they found a bulk flow of $188 ^{+119}_{-103} \kms$ towards \lb{290^{+39}_{-31}}{20^{+32}_{-32}} which also agrees well with our results. The remaining 425 high-z SNe show no significant bulk flow. This is expected since the peculiar velocity errors are typically $6\%$ of the distance to the source, and for this distant sample the errors per SNe measured in $\kms$ are extremely large.

Another interesting recent analysis of all peculiar velocities is by \citet{WatFelHud09}, who studied peculiar velocities mostly from Tully-Fisher, Fundamental Plane and SNe. They found that those subsamples had bulk flows consistent with each other\footnote{Except for the BCG sample of \citet{LauPos94}, which was excluded from further analysis.}.  They combined the individual peculiar velocity samples into a ``Composite'' sample of 4481 peculiar velocity tracers, which was found to have a MV50 bulk flow of 407 $\pm$ 81 $\kms$ towards \lberr{287}{9}{8}{6}. This result is inconsistent with $\Lambda$CDM at the 98\% CL.  However, their sample is not independent of ours. 103 of the 108 SNe which make up the ``Old'' subset of A1 are common to both A1 and Composite, although the latter takes SNe distances from \citet{TonSchBar03}.  When all SNe data are removed from the `Composite' data set, the two surveys become completely independent, and can be compared using the same formalism described in Section 3 of this paper and Section 5.1 of \citet{WatFelHud09}.  We find that the `Composite excluding SNe' MV50 bulk flow and the A1 MV50 results are consistent with each other, although the agreement is marginal:  $\chi^{2}$ of 6.4 for 3 directional degrees of freedom yields a 9 percent probability that the two results are consistent.

Lastly, these results can be compared directly to the expectations for a $\Lambda$CDM universe. A plot showing the expectations of the one dimensional rms for perfectly sampled Gaussian sphere can be found in the top three plots of figure 5 from \citet*{FelWatHud10}. The expectation for the one dimensional rms for a perfect Gaussian is 80 $\kms$. When you take into account the sparse sampling of the real A1 data set, this rises slightly to 91.2 $\kms$ assuming a $\sigma_{8}$ of 0.8 and $\Omega_{m}$ of 0.258.\footnote{In the rest of the paper because SN distances are insensitive to the value of $\Omega_{m}$ we used $\Omega_{m}$ = 0.3.  The predicted one dimensional rms drops slightly to 88.4 if this slightly higher value of $\Omega_{m}$ is used in the prediction}. If you then include the propagation of measurement uncertainties the total expected rms for surveys equivalent to ours at different locations ins space is 121 $\kms$. This prediction leads to $\chi^{2}$ of 3.70 for 3 directional degrees of freedom yielding a 70 percent probability that the A1 data set is consistent with $\Lambda$CDM.

\section{Predicted gravity field}

\subsection{Introduction}

The MV weighting scheme discussed above is designed to suppress the effects of small-scale flows that would otherwise ``alias'' power into the bulk-flow statistic. An alternate method for removing the effects of small-scale structure on flow measurements is to assume gravitational instability and linear perturbation theory Eq. \eqref{eq:vpred}  and to predict the peculiar velocities using a model of the density field (derived from an all-sky galaxy redshift survey). The result is a model-dependent correction to measured peculiar velocities which can separate local effects from large-scale density waves from outside the survey volume.  

Suppose we have an all-sky redshift survey that extends to a distance $R\sbr{max}$. We will model the peculiar velocity of a given SN located at position $\mathbf{r}$ by setting $v_{\mathrm{pred},i}$ of Eq. \eqref{eq:Chi2} to a function with two terms:
\begin{equation}
 \mathbf{v}_{\mathrm{pred}}(\mathbf{r}) = \frac{\beta}{4 \pi} \int_{0}^{R_{\mathrm{max}}} d^{3}\mathbf{r}'\delta_{g}(\mathbf{r}') \frac{\mathbf{r}'-\mathbf{r}}{\mid \mathbf{r}'-\mathbf{r}\mid^{3}} + \mathbf{U}
 \label{eq:pscz}
\end{equation} 
where $\beta = f / {b}$, $b$ is the linear bias between galaxy density and mass density, $\mathbf{U}$ is the \emph{residual} bulk flow of the volume being driven by mass structure beyond $R\sbr{max}$. In principle, the residual velocities have tidal (shear) and higher order terms. \citet{FelWatHud10} measured the the tidal and higher order terms for the ``Composite'' sample of \citet{WatFelHud09}, but found them to be small.  We neglect these terms here and model the residual as a simple bulk flow $\mathbf{U}$.  

The first term of $v_{\mathrm{pred}}$ is the predicted peculiar velocity induced by structure within the redshift survey volume $(r < R\sbr{max})$. The model is scaled by $\beta$ to match the observed peculiar velocities of the SNe tracers. The peculiar velocity data therefore yields information about $\Omega_{m}$ and $b$. The residual bulk flow $\mathbf{U}$ is the additional velocity of the entire redshift survey volume in the CMB reference frame, and is presumably due to sources beyond $R\sbr{max}$. In an ideal survey, $\mathbf{U}$ would be completely independent of any structure within $R\sbr{max}$. This de-coupling of $\mathbf{U}$ from $\beta$ means that $\mathbf{U}$ can be used to test consistency with $\Lambda$CDM + WMAP7 on large scales and $\beta$ can do so on smaller scales.

\subsection{Data and Method}

\begin{figure}
\begin{center}
\includegraphics[width=\columnwidth]{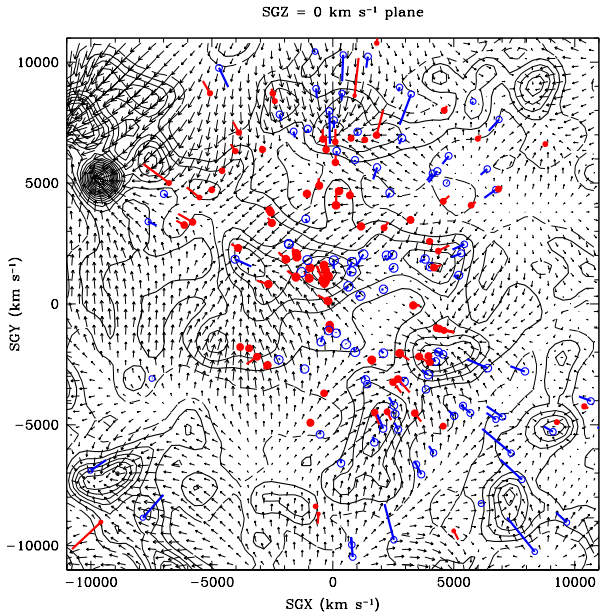}
\end{center}
\caption{The Supergalactic Plane. The PSCz galaxy density field is shown by the contours, predicted peculiar velocities as small black arrows, and measured supernova positions as ``tadpoles'' with dots showing measured positions and tails showing the magnitude of the measured radial peculiar velocity. The thick black contour corresponds to a $\delta =0$ (or contours where the density is the mean universal density). The red(filled circles) SNe have peculiar velocities away from the LG and the blue(open circles) SNe have peculiar velocities towards the LG.}\label{fig:Layers}
\end{figure}

The PSCz is both all-sky and deeper than, for example, the 2MRS \citep{SkrCutSti06}.   Here we use the PSCz density field reconstructed by \citet{BraTeoFre99}. For our study we applied the same 20,000 $\kms$ limit to the PSCz as we applied to the SNe.  The PSCz density field in the Supergalactic Plane is shown in  Fig. \ref{fig:Layers}. 

We fit the SNe data using the same method as in Section \ref{sec:MaxL}, but now with a new model as given by Eq. \eqref{eq:pscz}.  Since  the integral is specified by the PSCz density field, the free parameters are  $\beta$ and the three components of $\mathbf{U}$. Since the PSCz plus bulk flow is a better flow model than a simple bulk flow, we reduce the thermal component to 150 \kms, which is consistent with previous studies \citep{HudSmiLuc04}. 

\subsection{Results}

The results of the fits to each subset are given in Table \ref{tab:BetaMod}. We find that the results from independent subsets are consistent with each other. For the A1 sample, the magnitude of the residual bulk flow  was found to be 150 $\pm$ 43 $\kms$ in direction \lberr{345}{20}{8}{13}. This is significantly different from 0 at the $99.6\%$ CL.

\begin{figure}
\begin{center}
\includegraphics[width=\columnwidth]{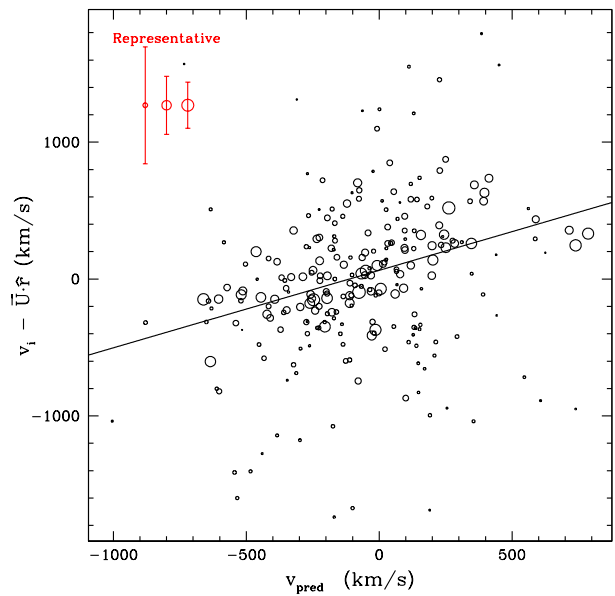}
\caption{The observed peculiar velocity minus the measured bulk flow
  as a function of the linear-theory-predicted peculiar velocity for
  each SN, assuming $\beta=1$. The circular symbol diameter scales
  with the inverse of the uncertainty (hence symbol area is
  proportional to weight). Representative error bars are shown in the
  top left. The slope is the fitted $\beta = 0.53$. }\label{fig:OtherBeta}
\end{center}
\end{figure}

The value of $\beta$ was found to be 0.53 $\pm$ 0.08, and is shown in  Fig. \ref{fig:OtherBeta}. The fit is sensitive to a single outlier, sn1992bh, for which the PSCz prediction is rather high (1719 km/s). Excluding this SN, we find $\beta$ = 0.57 $\pm$ 0.08.

Again we investigated the sensitivity of these results on the thermal noise term by changing it by $\pm$ 100 $\kms$. Again the magnitude off the flow only changed by $\pm$ 20 $\kms$, and the $\beta$ changed by $\pm$ 0.03. 

\begin{table*}
\caption{ Results from all three data subsets and the A1 full set with a linear perturbation theory model fit with a known matter distribution to fit $\beta$ and the residual flow $U$. Fit with $\beta$ as a free parameter and with a thermal noise of 150 $\kms$. Note the uncertainties quoted for the this method are the just propagated uncertainties from the measurements. }\label{tab:BetaMod}
\begin{tabular}{|l||c|         r@{$\pm$}l    |    r@{$\pm$}l    |    r@{$\pm$}l   |    r@{$\pm$}l    |     r@{$\pm$}l    |    r@{$\pm$}l    |  r@{$\pm$}l  |     c|     c    |} \hline\hline
               &$\#$ of SN&\multicolumn{2}{|c|}{Mag}&\multicolumn{2}{|c|}{l\arcdeg}&\multicolumn{2}{|c|}{b\arcdeg}&\multicolumn{2}{|c|}{U$_{X}$}&\multicolumn{2}{|c|}{U$_{Y}$}&\multicolumn{2}{|c|}{U$_{Z}$}&\multicolumn{2}{|c|}{$\beta$}&$\chi^{2}$&Dof\\		
\hline
\multicolumn{8}{|l|}{~}&\multicolumn{2}{|c|}{\kms}&\multicolumn{2}{|c|}{\kms}&\multicolumn{2}{|c|}{\kms}&\multicolumn{2}{|c|}{}& \\
\hline
Old            & 106  &190&59 &349&22&0 &14   &187& 60&-36&73 &0  &46   &.45&.11&139   &101\\		           
Hicken         & 113  &86 &77 &347&54&9 &41   &84 &87 &-19&83 &13 &66   &.62&.13&102   &108\\		           
Carnegie       & 26   &290&150&347&41&31&26   &240&170&-50&190&151&130  &.82&.33&21    &21 \\			  
First Amendment& 245  &150&43 &345&20& 8&13   &144&44 &-38&51 &20 &35   &.53&.08&270   &238\\ \hline\hline		 
\end{tabular}

\end{table*}

\subsection{Gravity field: cosmology and comparisons}

As noted above, the residual bulk flow $\mathbf{U}$ is significantly different from zero at the 99.6\% confidence level. This means that there are structures not found in the PSCz catalogue that contribute significantly to the total peculiar velocity of the LG. As discussed in detail in \citet{HudSmiLuc04} these structures could be missing from the PSCz because they are (i) outside the survey volume (ii) in the Zone Of Avoidance, or (iii) present but underrepresented. The latter scenario may arise because  the IRAS (far-infrared) selection on which PSCz is based is sensitive to dusty spirals, but less so to the mostly dust-free early types. \cite{LavHud11}, using the 6dF survey of 2MASS-selected galaxies, showed that the  Shapley and Horologium-Reticulum superclusters generate significantly more peculiar velocity than predicted by the PSCz, even allowing for a different $\beta$ for 2MASS galaxies.

We found that, for IRAS-selected galaxies, $\beta = 0.53 \pm 0.08$. This value of $\beta\sbr{I}$ is in good agreement with the IRAS average of $0.50 \pm 0.02$ reported by \citet{PikHud05},and with the SNIa-based result $\beta_{I} = 0.55\pm0.06$ of \cite{RadLucHud04}.  Comparing fitted values of $\beta$ between redshift surveys of different galaxy types is complicated by the fact that the bias factor $b$ need not be the same because different galaxy types may trace the underlining mass density differently.  This problem can be alleviated by noting that in linear theory the r.m.s.\ fluctuation of the survey galaxies, say in an 8 $\hmpc$ top hat sphere ($\sigma_{8,\mathrm{gal}}$) is proportional to the true matter r.m.s. fluctuations in a volume of the same size (i.e.\   $\sigma_{8,\mathrm{gal}} = b\sigma_{8,\mathrm{mass}}$).   Thus with our measured $\beta_{I}$ and the known $\sigma_{8,I}$ from the IRAS PSCz of $0.80\pm0.05$ \citep{HamTeg02}, we can calculate the degenerate parameter pair $f\sigma_{8}$ (where we dropped the subscript `mass'). Our value of $\beta$ corresponds to $f\sigma_{8} = 0.424 \pm 0.069$. 

We can then compare our $f\sigma_{8}$ to other studies.  \citet{DavNusMas11} compared the 2MRS density field and the SFI++ peculiar velocity data, and derived $f\sigma_{8} = 0.31\pm0.05$. This is lower than our result, but not significantly so ($1.5\sigma$). 

The $f\sigma_{8}$ parameter can also be derived from WMAP7 results. Recall that WMAP is observing fluctuations at an early epoch, when the perturbations were still well in the linear regime. To compare to WMAP7, we can convert our non-linear $\sigma_{8}$ into the equivalent linear value using the prescription of \citet{JusFelFry10}.  If we assume an $\Omega_{m}$ of 0.272, our $\sigma_{8,\mathrm{lin}}$ becomes 0.814 compared to its non-linear value of 0.867. Using this value of $\sigma_{8,\mathrm{lin}}$, $f\sigma_{8,\mathrm{lin}}$ drops to $0.40 \pm 0.07$, which is in excellent agreement with the results of WMAP7 \citep{LarDunHin11}: $f\sigma_{8,\mathrm{lin}}=0.39 \pm 0.04$.

\section{Discussion}

Attempts to determine the sources of the LG's motion amount to determining the factors in Eq. \eqref{eq:vpred}.  While early studies focussed on simple toy infall models, more recent studies have concentrated on models of the density field with the two free parameters $\beta$ and $\mathbf{U}$. For a single object, such as the LG itself, there is a trade-off between these parameters. Lower values of $\beta$ lead to larger values of $\mathbf{U}$, which are required in order to match the same $\mathbf{v}$ on the left-hand side of equation \ref{eq:pscz}. This degeneracy can be broken with more than one measurement. We have shown that the PCSz does not account for all of the motion of the LG, although it is plausible that some of the missing signal comes from within the survey volume in the form of extra infall into the highest-density superclusters.

An alternative explanation for the bulk flow has been proposed, namely that the CMB temperature dipole, or part thereof, is intrinsic and does not represent the peculiar velocity of the LG \citep[A ``tilted'' Universe:][]{Tur91,KasAtrKoc08b,MaGorFel11}.  This would lead to an illusory ``bulk flow'' which would extend well beyond the local volume, indeed to the horizon.  The apparent $1005\pm267 \kms$  bulk flow of $z < 0.25$ clusters claimed by \citet{KasAtrEbe10}, which is well outside the expectations of $\Lambda$CDM bulk flows, might be explained by such an effect.  In such a scenario, there is an additional ``bulk flow'' $\mathbf{U}\sbr{tilt}$ which never vanishes  no matter how deep a redshift survey $R\sbr{max}$ is used in Eq. \eqref{eq:pscz}.  Our measured $\mathbf{U}$ thus provides an upper limit on the $\mathbf{U}\sbr{tilt}$.  The amplitude of the bulk flow found by \citet{KasAtrEbe10} is inconsistent with our measurement of $\mathbf{U}=150\pm43$ $\kms$ .   However, amplitude of the \citet{KasAtrEbe10} bulk flow is systematically uncertain. If we compare only the direction of the A1 fit \lberr{345}{20}{8}{13}and the \citet{KasAtrEbe10} direction \lberr{296}{29}{39}{15}, the results are marginal:  they disagree at approximately the 90\% CL. Thus our results do not support the high amplitude bulk flow found by \citet{KasAtrEbe10}.

\section{Conclusion}

We have analyzed the peculiar velocities of a 245 SNe dataset dubbed  the ``First Amendment''. Overall, we have found that this new compilation is in marginal agreement with previous bulk flow results and is not in significant conflict with $\Lambda$CDM + WMAP7 predictions. The First Amendment compilation yields a bulk flow of 248 $\pm$ 87  $\kms$ in the direction \lberr{319}{25}{7}{13}. 

We have compared the peculiar velocities to the predictions from the IRAS PSCz and have found $\Omega\sbr{m}^{0.55}\sigma_{8,\mathrm{lin}}$ of $0.40 \pm 0.07$, which is in excellent agreement with the $\Lambda$CDM + WMAP7 predictions and other previous measurements.

A residual flow of 150 $\pm$ 43 $\kms$  \lberr{345}{20}{8}{13} was found for the IRAS Point Source Catalog as normalized with the First Amendment SNe. This may suggest that the IRAS PSCz undersamples massive dense superclusters such as the Shapley Concentration.  Nevertheless, the small amplitude of the residual flow is in conflict with ``tilted Universe'' scenarios such as might be favoured by the kSZ analysis of \citet{KasAtrEbe10}.

As its name suggests, the First Amendment compilation is readily extendible as new SNe are found and their distances are published. Ongoing surveys such as CfA4 (95 SNe, Hicken, private communication),  LOSS \citep{GanLiFil11}, Palomar Transit Factory \citep{LawKulDek09}, and CSP \citep[50 more distances expected soon]{StrPhiBol11}, and upcoming surveys such as SkyMapper \citep[100 SNe per year with $z < 0.085$,][]{KelSchBes07}, Pan-Starrs, and LSST will eventually  provide sufficient SNe to reduce the 20,000 $\kms$ bulk and residual flow uncertainties to the systematic limits.  Future results on $f\sigma_{8,\mathrm{lin}}$ are expected based on predicted peculiar velocities from the 2M++ redshift compilation \citep{LavHud11}. Additionally, although individually less precise, Fundamental Plane distances and peculiar velocities can contribute significant precision to bulk flow surveys by sheer numbers. We wish to re-analyze the full `Composite' data set from \citet{FelWat08} after replacing the 103 SNe currently contained in that data set with the 245 SNe of A1, as well as to add the Fundamental Plane peculiar velocities from NFPS \citep{SmiHudLuc06} and 6dF \citep{JonSauCol04} when they become available. For now the results are data-limited, but the future promises many fruitful results from many promising surveys, and we await them eagerly. 

\section{Acknowledgements}

We thank Alex Conley for helpful comparisons of SNe data, and acknowledge useful discussions with John Lucey, Roya Mohayaee and Jacques Colin. SJT and MJH acknowledge the financial support of NSERC. RPK acknowledges support from the NSF grant AST-0907903.

This research has made use of the NASA/IPAC Extragalactic Database (NED) which is operated by the Jet Propulsion Laboratory, California Institute of Technology, under contract with the National Aeronautics and Space Administration. 

\bibliographystyle{mn2e}
\bibliography{mjh}

\appendix

\section{Light curve parameter comparisons}

\begin{table*}
\caption{ Results for 162 SNe from \citet{HicWooBlo09} fit with the MLCS2k2 light curve fitter either with $R_{V}$ = 1.7 or $R_{V}$ = 3.1.}\label{tab:3p1vs1p7}
\begin{tabular}{|l||c|         r@{$\pm$}l    |    r@{$\pm$}l    |    r@{$\pm$}l   |    r@{$\pm$}l    |     r@{$\pm$}l    |    r@{$\pm$}l    |} \hline\hline
               &Number&\multicolumn{2}{|c|}{Mag}&\multicolumn{2}{|c|}{l\arcdeg}&\multicolumn{2}{|c|}{b\arcdeg}&\multicolumn{2}{|c|}{V$_{X}$}&\multicolumn{2}{|c|}{V$_{Y}$}&\multicolumn{2}{|c|}{V$_{Z}$}         \\
\hline
\multicolumn{8}{|l|}{ML Thermal noise=250 \kms}&\multicolumn{2}{|c|}{\kms}&\multicolumn{2}{|c|}{\kms}&\multicolumn{2}{|c|}{\kms} \\
\hline
$R_{V}$ = 1.7  &162       &220&70 &298&18&9  &14&103&68 &-191&73 & 35& 52\\
$R_{V}$ = 3.1  &162       &175& 70&310& 25&14&18 &108&70 &-131& 75& 43& 53\\
 \hline
 \hline
\end{tabular}

\end{table*}

The A1 data set is composed of three different SNe catalogues; this complicates the description of the light curve fitting procedures used because the catalogues used different methods.  In the `Old' sample are 34 SNe  from \citet{JhaRieKir07}; most of which are fit using the MLCS2k2 light curve fitter with a reddening law parameter $R_{V}$ of 3.1 (for SN with high extinction, $R_{V}$ was a free fit parameter with a tight prior of 3.1). The remaining 72 SN in `Old' are from \citet{HicWooBlo09} and are also fit using MLCS2k2, but with a reddening law parameter $R_{V}$ of 1.7. The second data set, which we refer to as `Hicken', contains the remaining 113 SNe from \citet{HicWooBlo09} and are all fit with MLCS2k2 with a reddening law parameter $R_{V}$ of 1.7. The `Carnegie' set containing 28 SNe were fit with a $R_{V}$ as a free variable. The light curve fitter used for the `Carnegie' set is described in detail in the original paper \citet{SauSutMad00}.

For SNe fit by \citet{HicWooBlo09}, four distances were reported for each SN. We use the distances reported using the MLCS2k2 fitting procedure rather than either of the SALT procedures for multiple reason. To start the MLCS2k2 process determines host reddening on a case-by-case basis. Furthermore, of the two published MLCS2k2 methods we use the results with a reddening law parameter $R_{V}$ of 1.7 instead of 3.1 since \citet{HicWooBlo09} show that the Hubble residuals for high-extinction SNIa's using  $R_{V}$=3.1 are systematically negative, (suggesting that the extinction is overestimated).  We study the effect the choice of $R_{V}$ parameter has upon bulk flow measurements to explore systematics. \citet{HicWooBlo09} provide distances to 162 SNe using both $R_{V}$ = 1.7 and $R_{V}$ = 3.1. 
We fit both of these data sets for bulk flows using the ML method to investigate the systematics. The results of this comparison can be seen in table \ref{tab:3p1vs1p7}.  Although the results for the two light curve fitters agree to less than one $\sigma$ in each of the 3 degrees of freedom, the data sets are fit to the same light curves, so they are not independent. This result highlights how large the systematic errors are for bulk flow surveys, in part reflected by the large $\sigma_{SN}$, which in most cases dominates the uncertainty budget for peculiar velocity surveys.

\label{lastpage}

\end{document}